\documentclass[preprint,aps,eqsecnum,superscriptaddress]{revtex4}
\usepackage{graphicx,amsfonts}
\usepackage{amsmath}
\usepackage{slashed}
\newcommand{\be}{\begin{equation}}
\newcommand{\ee}{\end{equation}}
\newcommand{\bea}{\begin{eqnarray}}
\newcommand{\eea}{\end{eqnarray}}
\newcommand{\bml}{\begin{subequations}}
\newcommand{\eml}{\end{subequations}}
\newcommand{\pa}{\partial}

\newcommand{\vx}{\vec{x}}

\newcommand{\vk}{\vec{k}}

\newcommand{\e}{\epsilon}
\newcommand{\ve}{\varepsilon}

\begin{document}

\title {On the quantum dynamics of non-commutative systems}

\author{F. S. Bemfica}
\author{H. O. Girotti}
\affiliation{Instituto de F\'{\i}sica, Universidade Federal do Rio Grande
do Sul, Caixa Postal 15051, 91501-970 - Porto Alegre, RS, Brazil}
\email{fbemfica, hgirotti@if.ufrgs.br}

\begin{abstract}
This is a review paper concerned with the global consistency of the quantum dynamics of non-commutative systems. Our point of departure is the theory of constrained systems, since it provides a unified description of the classical and quantum dynamics for the models under investigation. We then elaborate on recently reported results concerned with the sufficient conditions for the existence of the Born series and unitarity and turn, afterwards, into analyzing the functional quantization of non-commutative systems. The compatibility between the operator and the functional approaches is established in full generality. The intricacies arising in connection with the explicit computation of path integrals, for the systems under scrutiny, is illustrated by presenting the detailed calculation of the Feynman kernel for the non-commutative two dimensional harmonic oscillator.

\vspace{1.0cm}
\end{abstract}

\maketitle
\newpage

\section{Introduction}
\label{sec:level1}

In this work we shall be concerned with quantum systems whose dynamics is described by a self-adjoint Hamiltonian $H(Q,P)$ made up of the Cartesian coordinates $Q^l, l= 1,\ldots, N$ and their canonically conjugate momenta $P^j, j = 1,\ldots, N$\cite{footnote 0}. However, unlike the usual case, coordinates and momenta are supposed to obey the non-canonical equal-time commutation rules

\bml
\label{1}
\begin{eqnarray}
&&\left[Q^l, Q^j\right] = -2 i\hbar\theta^{lj},\label{mlett:a1}\\
&&\left[Q^l, P_j\right] = i\,\hbar\,\delta^{l}_{\,\,j},\label{mlett:b1}\\
&&\left[P_l, P_j\right] = 0.\label{mlett:c1}
\end{eqnarray}
\eml

\noindent
The distinctive feature is, of course, that the coordinate operators do not commute among themselves. The lack of non-commutativity of the coordinates is parameterized by the real antisymmetric $N \times N$ constant matrix $\|\theta\|$\cite{footnote1}. In Refs.\cite{Chaichian1,Gamboa1,Gamboa2,Girotti1,Bemfica} the reader will find specific examples of noncommutative systems whose quantization has been successfully carried out. While Ref.\cite{Chaichian1} is concerned with the distortion provoked by the non-commutativity on the spectrum of the hydrogen atom, Refs.\cite{Gamboa1,Gamboa2,Girotti1} deal with the noncommutative two-dimensional harmonic oscillator, an exactly solvable model. In Ref.\cite{Bemfica} the authors elaborate about the effects of the non-commutativity in the case of a multi-particle system: the electron gas.

However, the question on whether non-commutative mechanics is, on general grounds, a sound quantum theory remains open. The answer for this question calls for model independent developments. Our purpose in this work is to summarize and discuss the key contributions that have been made in this respect\cite{footnote101}.

We shall first focus on the implementation of the classical-quantum transition for non-commutative systems. To the best of our knowledge, only the theory of constrained systems furnishes the appropriate tools for this purpose. As a by product, we shall also verify that non-commutativity always amounts to non-local interactions. This is our Section 2.

Unitarity is at the heart of any quantum theory since it secures probability conservation. For non-commutative models, a throughout investigation of this property is presented in Section 3.

In Section 4 we take, once more, advantage of the correspondence existing between non-commutative models and constrained systems to formulate the non-commutative quantum dynamics in terms of path integrals. We also verify here the compatibility of the operator and functional approaches.

Section 5 is dedicated to illustrate about the difficulties encountered when explicitly computing functional integrals for non-commutative systems. Being forced by didactics to work out an exactly solvable model, we restrict ourselves to consider the problem of finding the Feynman kernel for the non-commutative two-dimensional harmonic oscillator.

The conclusions and final remarks are contained in Section 6.

\section{Classical-quantum transition for noncommutative systems}
\label{sec:level2}

To start with, we notice that the classical counterpart of a quantum system involving non-commuting coordinates must be a constrained system\cite{footnote2}. Indeed, the equal time algebra in Eq.(\ref{mlett:a1}) could not have been abstracted from a Poisson bracket algebra, simply because the Poisson bracket of two coordinates vanishes.

Now, the problem of finding a constrained system mapping onto the noncommutative theory specified in (\ref{1}) has already been solved\cite{Deriglazov1}. Its classical dynamics is described by the Lagrangian\cite{footnote3}

\be
\label{2}
 L\,=\,v_j\,{\dot q}^j\,-\,h_0(q^j , v_j)\,+\,{\dot v}_j\,\theta^{jl}\,v_l\,,
\ee

\noindent
where repeated indices are summed from 1 to $N$. The constraint structure of this system reduces to the primary second-class constraints $G_i\,\equiv\,p_i\,-\,v_i\,\approx\, 0\,,\,T^i\,\equiv\,\pi^i\,-\,\theta^{ij} v_j\,\approx\,0\,$, where $p_i$ ($\pi^i$) is the momentum canonically conjugate to the generalized coordinate $q^i$ ($v_i$) and the sign of weak equality ($\approx$) is being used in the sense of Dirac\cite{Dirac1}. As for the canonical Hamiltonian, one finds that

\be
\label{4}
h(q , p)\,=\,h_0(q , p)\,.
\ee

\noindent
It may also be checked that the Faddeev-Popov matrix turns out to be unimodular and constant. Then, the computation of the Dirac brackets (DB) yields

\bml
\label{5}
\bea
&&[q^j, q^k]_{DB} = -2 \theta^{lk},\label{mlett:a5}\\
&&[q^j, p_k]_{DB} = \delta^{j}_{\,\,k},\label{mlett:b5}\\
&&[p_j, p_k]_{DB} = 0\label{mlett:c5}.
\eea
\eml

We do not need to compute explicitly the DB's involving $v^{\prime}$s and/or $\pi^{\prime}$s since, by definition\cite{Dirac1,Fradkin1,Sundermeyer1,Girotti2,Gitman1,Sudarshan1}, within the DB algebra the constraints hold as strong identities. In fact, at this stage of the formulation we may eliminate the variables $v$ and $\pi$ in favor of $q$ and $p$. However, $q$ and $p$ may not be referred to as the physical phase space variables because their DB's differ from the corresponding Poisson brackets (PB)\cite{Dirac1,Fradkin1,Sundermeyer1,Girotti2,Gitman1,Sudarshan1}. Presently, the construction of the physical phase space variables, ($x$ , $k$), in terms of $q$ and $p$ is straightforward. Indeed, one may easily verify that

\bml
\label{7}
\bea
&&x^j\,\equiv\,q^j\,-\,\theta^{j l}\,p_l\,,\label{mlett:a7}\\
&&k_j\,\equiv\,p_j\,\label{mlett:b7}
\eea
\eml

\noindent
and (\ref{5}) lead to $[\xi^j, \xi^{ l}]_{DB}\,=\,[\xi^j, \xi^{ l}]_{PB}$, for $\xi$ either $x$ or $k$. All that remains to be done to erase any remaining trace of the constraints is to rewrite the Hamiltonian in (\ref{4}) in terms of the physical variables, namely,

\be
\label{9}
h\left(q , p \right)\,\equiv\,h\left(x^j + \theta^{jk}k_{\,\,k}\,,\,k_{\,\,j} \right)\,.
\ee

\noindent
One may confirm that the Hamiltonian equations of motion for the physical variables possess the canonical form.

We turn next into quantizing the classical model described above. Within the operator framework the quantization is implemented by first promoting $q$ and $p$ into self-adjoint operators, $Q$ and $P$, respectively. The classical-quantum correspondence rule demands that they verify the equal-time commutator algebra abstracted from the corresponding DB's. Furthermore, up to ordering ambiguities, the Hamiltonian operator $H(Q , P)$ can be read off from $h(q , p)$ given at Eq.(\ref{4}). It is then clear that the noncommutative system (\ref{1}) is the quantized version of the classical constrained system defined in (\ref{2}).

The quantization procedure also requires the finding of a realization of the algebra (\ref{1}) in terms of matrices, i.e., of a representation. The fact that the coordinates do not commute among themselves rules out the possibility for the existence of a set of common $Q$-eigenvectors. However, the self-adjoint operators $X$ and $K$, which arise from the classical-quantum correspondence $x \longrightarrow X$, $k \longrightarrow K$, obey, by definition, the algebra abstracted from the corresponding PB's, i.e.,

\bml
\label{II-12}
\begin{eqnarray}
&&\left[X^{l}, X^{j}\right]\, = \,0\,,\label{mlett:aII-12}\\
&&\left[X^{l}, K_{j}\right] = i\,\hbar\,\delta^{l}_{\,\,j}\,,\label{mlett:bII-12}\\
&&\left[K_{l}, K_{j}\right] = 0\,.\label{mlett:cII-12}
\end{eqnarray}
\eml

\noindent
Hence, the common $X$-eigenvectors ($|{\vec x}> \equiv | x^{1},\ldots,x^{l},\ldots,x^{N}>$) provide a basis in the space of states for representing the algebra (\ref{1}).

For a Hamiltonian

\be
\label{II-14}
H(Q ,P) = \frac{P_l P_l}{2 M} + V(Q)
\ee

\noindent
and, therefore,

\be
\label{II-15}
H(X^{l}\, + \,\theta^{lj}\, K_{j}\,,\, K_{l}) = \frac{K_l K_l}{2 M} + V(X^{l} + \theta^{lk}\, K_k)\,,
\ee

\noindent
it has been shown elsewhere\cite{Gamboa1,Gamboa2,Girotti1} that the time evolution of the system, in the Schr\"odinger picture, is described by the wave equation

\be
\label{II-16}
- \frac{\hbar^2}{2M} \nabla_{x}^2 \Psi(x,
t) + V (x)\star \Psi(x, t) = i \hbar \frac{\pa
 \Psi(x, t)}{\pa t}\,,
\ee

\noindent
where $\nabla_{x}^2$ designates the $N$th-dimensional Laplacian, $M$ is a constant with dimensions of mass while $\star$ denotes the Gr\"onewold-Moyal product\cite{Gronewold,Moyal,Filk1}, namely,

\bea
\label{II-17}
V (x)\star \Psi(x, t)&\equiv& V(x) \left[\exp \left(- i
\hbar \overleftarrow{\frac{\pa}{\pa
x^l}} \,\theta^{lj}\, \overrightarrow{\frac{\pa}{\pa
x^{j}}}\right)\right] \Psi(x, t)\nonumber\\
&=&\,V \left(x^{j}\,-\,i \hbar \,\theta^{jl}\,\frac{\partial}{\partial x^{l}}\right)\Psi(x, t)\,.
\eea

\noindent
It is worth mentioning that in Refs.\cite{Chaichian1,Gamboa1,Gamboa2,Girotti1} Eq.(\ref{II-16}) has been solved for some specific models.

The non-local nature of the right hand side of Eq.(\ref{II-17}) should be noticed. Hence, as stated in Section 1, the quantized version of non-commutative systems always involve non-local interactions.

\section{Born series and Unitarity in noncommutative quantum mechanics}
\label{sec:level3}

Unitarity is of paramount importance for the consistency of a quantum theory. Presently, the non locality of the interaction casts doubts on whether the self-adjointness of the Hamiltonian suffices, by itself, to render the scattering operator $S$ unitary. A two steps procedure can be adopted to clarify this issue. First, one proves that, under certain restrictions to be imposed on the potential ($V$), there exists a convergent Born series expansion for the matrix elements of the transition operator $T$ ($S \equiv I - 2 \pi i T$).  For $V = g U$, with $g$ a dimensionless coupling constant, the Born series becomes a power series expansion in $g$. Then, unitarity will be shown to hold order by order in $g$\cite{Bemfica4}.

\subsection{Born series}

Let us return, for a while, to commutative quantum mechanics and consider a system whose dynamics is described by the self-adjoint Hamiltonian operator

\be
\label{III-1}
H\,=\,H_0\,+\,V\left(X\right)\,,
\ee

\noindent
where $H_0 \equiv K_lK_l/2M$ will be referred to as the free Hamiltonian. Notice that $H = H^{\dagger}$ enforces $V = V^{\dagger}$ since the kinetic energy part $H_0$ is, by construction, self-adjoint. From inspection follows that $H_0$ does not possess bound states and its continuum energy spectrum is characterized by $E > 0$. By assumption, the same applies for the continuum spectrum of $H$ although this operator may also possess bound states. Furthermore, we shall keep everywhere in this Section $\hbar = 1$.

For the quantum system under consideration all observables can be obtained from the operator $T(W)$ defined by the integral equation

\be
\label{III-2}
T(W)\,=\,V\,+\,V\,G_0^{(+)}(W)\,T(W)\,,
\ee

\noindent
where $G_0^{(+)}(W)\,=\,\left[W\,-\,H_0\,+\,i\ve\right]^{- 1}$ is the free Green function for outgoing boundary conditions. By iterating the right hand side of Eq.(\ref{III-2}) one obtains $T$ as a series,

\be
\label{III-4}
T(W)\,=\,V\,+\,V\,G_0^{(+)}(W)\,V\,+\, V\,G_0^{(+)}(W)\,V\,G_0^{(+)}(W)\,V \,+\,\cdots\,,
\ee

\noindent
known as the Born series.

The problem of determining the necessary and sufficient conditions for the Born series to converge was solved by Weinberg\cite{Weinberg} long ago. He considers the eigenvalue problem

\be
\label{III-5}
G_0^{(+)}(W)\,V\,|\psi_{\nu}(W)\rangle\,=\,\eta_{\nu}(W)\,|\psi_{\nu}(W)\rangle\,.
\ee

\noindent
Since the operator $G_0^{(+)}(W)\,V$ is not Hermitean, the eigenvalues $\eta(W)$ may be complex. As for the eigenstates, $|\psi_{\nu}(W)\rangle$, they are assumed to be of finite norm. $W$ is kept negative or complex and is allowed to approach the positive real axis from above. From Eqs.(\ref{III-4}) and (\ref{III-5}) one obtains

\be
\label{III-6}
T(W)\,|\psi_{\nu}(W)\rangle\,=\,\left[\sum_{n = 0}^{\infty} \eta^n_{\nu}(W) \right]\,V\,|\psi_{\nu}(W)\rangle\,.
\ee

\noindent
It was demonstrated by Weinberg\cite{Weinberg} that

\be
\label{III-7}
|\eta_{\nu}(W)|\,<\,1\,,\qquad \forall \,\nu\,,
\ee

\noindent
is a necessary and sufficient condition for the Born series to converge.

We now want to solve the analogous problem for noncommutative quantum mechanics, the essential difference from above being that instead of $V = V(X)$ we have $V = V(X^l + \theta^{lj}\,K_j)$. As point of departure, we start by invoking (\ref{III-5}) to cast Eq.(\ref{III-7}) as

\bea
\label{III-8}
&&\frac{|\langle {\vec k}\,|G_0^{(+)}(W)\,V\,|\psi_{\nu}(W)\rangle|}{|\langle {\vec k}\,|\psi_{\nu}(W)\rangle|}\,=\,\frac{1}{|W - \frac{{\vec k}^2}{2 M}\,+\,i \ve|}\,\frac{|\langle {\vec k}\,|\,V\,|\psi_{\nu}(W)\rangle|}{|\langle {\vec k}\,|\psi_{\nu}(W)\rangle|}\nonumber\\
&&\,=\,\frac{1}{|W - \frac{{\vec k}^2}{2 M}\,+\,i \ve|}\,\frac{1}{|\langle {\vec k}\,|\psi_{\nu}(W)\rangle|}\bigg|\int d^Nk^{\prime}\,\langle {\vec k}\,|\,V\,|{\vec k}^{\,\prime}\rangle \langle{\vec k}^{\,\prime}|\psi_{\nu}(W)\rangle\bigg|\nonumber\\
&&<\,1\,,\qquad \forall \,\nu\,.
\eea

\noindent
Let us concentrate on the linear momentum integral in the right hand side of Eq.(\ref{III-8}). Since

\be
\label{III-9}
\bigg|\int d^Nk^{\prime}\,\langle {\vec k}\,|\,V\,|{\vec k}^{\,\prime}\rangle \langle{\vec k}^{\,\prime}|\psi_{\nu}(W)\rangle\bigg| \,\leq\,\int d^Nk^{\prime}\,\big|\langle {\vec k}\,|\,V\,|{\vec k}^{\,\prime}\rangle \langle{\vec k}^{\,\prime}|\psi_{\nu}(W)\rangle\big|\,,
\ee

\noindent
one concludes that

\bea
\label{III-10}
\frac{1}{|W - \frac{{\vec k}^2}{2 M}\,+\,i \ve|}\,\frac{1}{|\langle {\vec k}\,|\psi_{\nu}(W)\rangle|}\,\int d^Nk^{\prime}\,\big|\langle {\vec k}\,|\,V\,|{\vec k}^{\,\prime}\rangle \big| \,\big|\, \langle{\vec k}^{\,\prime}|\psi_{\nu}(W)\rangle\big|\,<
\,1\,\qquad \forall \,\nu
\eea

\noindent
is a sufficient although not necessary condition for the convergence of the Born series. In other words, (\ref{III-10}) selects a subset of potentials for which the Born series certainly converge.

To proceed further on we shall be needing $\big|\langle {\vec k}\,|\,V\,|{\vec k}^{\,\prime}\rangle\big|$. Then, we start by looking for

\bea
\label{III-11}
&&\langle {\vec k}\,|\,V(X^l + \theta^{lj}\,K_j)\,|{\vec k}^{\,\prime}\rangle\,=\,\int d^Nx\,\phi_{\vk}^{\star}(\vx)\,V\left(x^l\,-\,i\,\,\theta^{lj}\,\frac{\pa}{\pa x^j}\right)\,\phi_{{\vk}^{\,\prime}}(\vx)\nonumber\\
&&=\,\int d^Nx\,\phi_{\vk}^{\star}(\vx)\,\left[V(\vx) \star \phi_{{\vk}^{\,\prime}}(\vx)\right]\,=\,\int d^Nx\,\phi_{\vk}^{\star}(\vx) \star V(\vx) \star \phi_{{\vk}^{\,\prime}}(\vx)\nonumber\\
&&=\,\int d^Nx\, V(\vx)\,\left[\phi_{{\vk}^{\,\prime}}(\vx) \star \phi_{\vk}^{\star}(\vx) \right]\,,
\eea

\noindent
where
\be
\label{III-12}
\phi_{{\vk}}(\vx) = \frac{1}{(2 \pi )^{\frac{N}{2}}}\,e^{i\,k_j x^j}\,,
\ee

\noindent
is the eigenfunction of the linear momentum ${\vec K}$, corresponding to the eigenvalue ${\vk}$. By invoking Eq.(\ref{II-17}) one, then, finds

\bea
\label{III-13}
\phi_{{\vk}^{\,\prime}}(\vx) \star \phi_{\vk}^{\star}(\vx)\,=\,\phi_{{\vk}^{\,\prime}}(\vx) \,\left[\exp \left(- i
 \overleftarrow{\frac{\pa}{\pa
x^l}} \,\theta^{lj}\, \overrightarrow{\frac{\pa}{\pa
x^{j}}}\right)\right]\,\phi_{\vk}^{\star}(\vx)\,=\,e^{-\,i\,{\vec k}^{\,\prime}\,\wedge\,\vk}\,\phi_{{\vk}^{\,\prime}}(\vx)\, \phi_{\vk}^{\star}(\vx)\,,
\eea

\noindent
where

\be
\label{III-131}
{\vec k}^{\,\prime}\,\wedge\,\vk\,\equiv\,k_l^{\prime}\,\theta^{l j}\,k_j\,.
\ee

\noindent
Clearly, Eqs.(\ref{III-13}) and (\ref{III-11}) amount to

\be
\label{III-14}
\langle {\vec k}\,|\,V(X^l + \theta^{lj}\,K_j)\,|{\vec k}^{\,\prime}\rangle\,=\,e^{-\,i\,{\vec k}^{\,\prime}\,\wedge\,\vk}\,\langle {\vec k}\,|\,V(X^l)\,|{\vec k}^{\,\prime}\rangle\,
\ee

\noindent
and, as consequence,

\be
\label{III-15}
\big|\langle {\vec k}\,|\,V(X^l + \theta^{lj}\,K_j)\,|{\vec k}^{\,\prime}\rangle\big|\,=\, \big|\langle {\vec k}\,|\,V(X^l)\,|{\vec k}^{\,\prime}\rangle \big|\,.
\ee

This result connects the commutative with the noncommutative regimes. Therefore, if $V(X)$ verifies Eq.(\ref{III-10}) so does $V(X^l + \theta^{lj}\,K_j)$ or, what amounts to the samething, for the restricted subclass of potentials verifying Eq.(\ref{III-10}) the convergence of the Born series holds for both, the commutative and the noncommutative versions of the model.

\subsection{Unitarity in noncommutative quantum mechanics}

The scattering amplitude $f( {\vk}^{\,\prime}\,,\,\vk)$ is given in terms of the $T$-matrix by

\be
\label{III-16}
f( {\vk}^{\,\prime}\,,\,\vk)\,\equiv\,-\,4 \pi^2 M\,T( {\vk}^{\,\prime}\,,\,\vk)\,,
\ee

\noindent
where $T( {\vk}^{\,\prime}\,,\,\vk)$ is short for $\langle {\vk}^{\,\prime}|T|\vk \rangle$. Unitarity demands that

\be
\label{III-17}
\Im\,f( {\vk}\,,\,\vk)\,=\,\frac{k}{4 \pi}\,\int\,d\Omega_{\vk^{\,'}}\,\big|f( {\vk}^{\,'}\,,\,\vk)\big|^2\,,
\ee

\noindent
where $k = |\vk|$ and $d\Omega_{\vk^{\,'}}$ is the element of solid angle centered around $\vk^{\,'}$.

Our purpose here is to check (\ref{III-17}) by taking advantage of the Born series representation for $T$. It will be assumed that the potential $V$ contains a dimensionless coupling constant ($g$) that enables one to write $V = g\, U$. Then, the Born series in Eq.(\ref{III-4}) becomes a power series in $g$. Correspondingly, Eq.(\ref{III-17}) translates into

\be
\label{III-19}
\frac{4\,\pi}{k}\,\Im\,f^{(n)} ({\vk}\,,\,\vk)\,=\,\int\,d\Omega_{\vk^{\,'}}\,\sum_{i = 1}^n\,f^{(i)^{\star}}( {\vk}^{\,'}\,,\,\vk)\,f^{(n\,-\,i)}( {\vk}^{\,'}\,,\,\vk)\,,
\ee

\noindent
where $n$ is a positive integer,

\be
\label{III-191}
f^{(n)}( {\vk}^{\,'}\,,\,\vk)\,=\,-\,4 \pi^2 M\,T^{(n)}(\vk\,' , \vk)\,,
\ee

\noindent
and

\bea
\label{III-192}
T^{(n)}(\vk , \vk\,')\,=\,\langle \vk\, \big|\overbrace{V G_0^{(+)}(E)V\cdots V G_0^{(+)}(E)V}^{n\, \mbox{factors}\, V; (n - 1)\, \mbox{factors}\,G_0^{(+)}(E)}  \big|\vk\,'\rangle\,.
\eea

Let us first analyze the contributions to the scattering amplitude for $n = 1$. Clearly, the right hand side in (\ref{III-19}) does not contain terms of order $g^1$. Then, no term of order $g^1$ should arise in $\Im\,f^{(1)}( {\vk\,,\,\vk})$. We know that this is the case in the commutative version of the theory, since the hermiticity of $V$ secures $\Im \,\langle {\vec k}\,|\,V(X^l)\,|{\vec k}\rangle\,=\,0$. As for the noncommutative case, we observe that for $\vk^{\,\prime}\,=\,\vk$ (forward direction) the exponent in the right hand side of (\ref{III-14}) vanishes and, therefore, $\Im \,\langle {\vec k}\,|\,V(X^l + \theta^{lj}\,K_j)\,|{\vec k}\rangle\,=\,\Im \,\langle {\vec k}\,|\,V(X^l)\,|{\vec k}\rangle\,=\,0$, as required.

To verify Eq.(\ref{III-19}) for arbitrary $n$ we start by claiming that

\bea
\label{III-20}
&&\Im\,\int\,d^Nk^{\prime}\,\frac{T^{(m)^{\star}}(\vk\,' , \vk)\,T^{(p)}(\vk\,' , \vk)}{\frac{k^2}{2M}\,-\,\frac{k^{'\,2}}{2M}\,+\,i\ve}\,=\,
\Im\,\int\,d^Nk^{\prime}\,\frac{T^{(m + 1)^{\star}}(\vk\,' , \vk)\,T^{(p - \,1)}(\vk\,' , \vk)}{\frac{k^2}{2M}\,-\,\frac{k^{'\,2}}{2M}\,+\,i\ve}\nonumber\\
&&-\,M\,k\,\pi\,\int\,d\Omega_{\vk\,'}\,\left[ T^{(m)^{\star}}(\vk\,' , \vk)\,T^{(p)}(\vk\,' , \vk)\,+\,T^{(p)^{\star}}(\vk\,' , \vk)\,T^{(m)}(\vk\,' , \vk)\right]\,,
\eea

\noindent
whose proof is straightforward but will be omitted for reasons of space. Then, consider

\bea
\label{III-21}
&&\Im T^{(n)}(\vk , \vk)\,=\,\Im \int\,d^Nk^{\prime}\,\frac{T^{(1)^{\star}}(\vk\,' , \vk)\,T^{(n - 1)}(\vk\,' , \vk)}{\frac{k^2}{2M}\,-\,\frac{k^{'\,2}}{2M}\,+\,i\ve}\nonumber\\
&&=\,\Im\,\int\,d^Nk^{\prime}\,\frac{T^{(2)^{\star}}(\vk\,' , \vk)\,T^{(n - \,2)}(\vk\,' , \vk)}{\frac{k^2}{2M}\,-\,\frac{k^{'\,2}}{2M}\,+\,i\ve}\nonumber\\
&&-\,M\,k\,\pi\,\int\,d\Omega_{\vk\,'}\,\left[ T^{(1)^{\star}}(\vk\,' , \vk)\,T^{(n - \,1)}(\vk\,' , \vk)\,+\,T^{(n - \,1)^{\star}}(\vk\,' , \vk)\,T^{(1)}(\vk\,' , \vk)\right]\,,
\eea

\noindent
where in going from the second to the third term of the equality we have used (\ref{III-20}) for $ m = 1$ and $p = n - 1$. It is not difficult to see that by applying this procedure $ (n - 2)$ times one ends up with

\bea
\label{III-22}
&&\Im T^{(n)}(\vk , \vk)\,=\,\Im T^{(n)^{\star}}(\vk , \vk)\nonumber\\
&&- 2 M k \pi\,\int\,d\Omega_{\vk\,'}\,\left[ T^{(1)^{\star}}(\vk\,' , \vk) T^{(n - \,1)}(\vk\,' , \vk) + \cdots + T^{(n - \,1)^{\star}}(\vk\,' , \vk) T^{(1)}(\vk\,' , \vk)\right]\,,
\eea

\noindent
which, after recalling that $\Im T^{(n)^{\star}}(\vk , \vk)\,=\,-\,\Im T^{(n)}(\vk , \vk)$,  goes into

\bea
\label{III-23}
\Im T^{(n)}(\vk , \vk)\,=\,-\,M \,k \,\pi\,\int\,d\Omega_{\vk\,'}\,\sum_{i = 1}^n\,T^{(i)^{\star}}( {\vk}^{\,'}\,,\,\vk)\,T^{(n\,-\,i)}( {\vk}^{\,'}\,,\,\vk)\,.
\eea

\noindent
This last equation reproduces Eq.(\ref{III-19}) in terms of $T$-matrix elements and, hence, concludes the purported proof of unitarity. It applies equally well for the commutative and the noncommutative cases.

\section{The functional formulation of the quantum dynamics of noncommutative systems}
\label{sec:level4}

 In this Section we develop the functional formulation of the quantum dynamics of noncommutative systems. To reach this goal we shall take advantage of the equivalence described in Section 2, since the functional formulation of the dynamics of constrained systems is, by now, a well known theoretical tool. In fact, we have already at hand all the ingredients entering the phase space path integral defining the generating functional of Green functions ($Z[J , S]$), which reads\cite{Fradkin1}

\bea
\label{IV-1}
&&Z[J , S]\,=\,{\mathcal C}\,\int \left[{\mathcal D}q\right] \int \left[{\mathcal D}v\right]
\int \left[{\mathcal D}p\right]\int \left[{\mathcal D}\pi\right]\,\left\{\prod_{j = 1}^{N}\,\delta[p_j - v_j]\right\}\nonumber\\
&\times&\left\{\prod_{j = 1}^{N} \delta[\pi^j - \theta^{jk} v_k]\right\}\,\exp\left\{\frac{i}{\hbar} \int_{t_{in}}^{t_f} dt\left[ p_j\, {\dot q}^j\,+\,\pi^j\, {\dot v}_j\,-\,h(q , p)\,\right. \right.\nonumber\\
&+&\left. q^j\, J_j\,+\,p_j\, S^j\right]\bigg\}\,.
\eea

\noindent
Here, $J$ and $S$ are external sources for $q$ and $p$, respectively, while ${\mathcal C}$ is a normalization constant to be chosen such that $Z[J = 0 , S = 0] = 1$. After performing the functional integrals on $\pi$ and $v$ one ends up with

\bea
\label{IV-2}
&&Z[J , S]\,=\,{\mathcal C}\,\int \left[{\mathcal D}q\right]
\int \left[{\mathcal D}p\right]\nonumber\\
&\times&\exp\left\{\frac{i}{\hbar} \int_{t_{in}}^{t_f} dt\left[ p_j\, {\dot q}^j\,-\,p_j\,\theta^{jk} {\dot p}_k\,-\,h(q , p)\,
+ q^j\, J_j\,+\,p_j\, S^j\right]\right\}\,.
\eea

Thus far we have succeeded in eliminating all the redundant degrees of freedom and, therefore, in expressing $Z[J , S]$ as a phase space path integral over independent variables.  However, this is not the end of the story because, as we already pointed out, $q$ and $p$ are not canonical phase space variables. On the other hand, a proof of existence for $Z[J , S]$ written as a phase space path integral over independent canonical variables ($x$, $k$) was obtained by Fradkin and Vilkovisky\cite{Fradkin1}. Presently, we find such expression by performing the non-canonical transformation (\ref{7}) which, in turns, allows us to cast Eq.(\ref{IV-2}) as

\bea
\label{IV-6}
&&Z[J , S|x_f,t_f;x_{in},t_{in}]\,=\,{\mathcal C}\,\int \left[{\mathcal D}x\right]
\int \left[{\mathcal D}k\right]\nonumber\\
&\times&\exp\left\{\frac{i}{\hbar} \int_{t_{in}}^{t_f} dt\left[ k_j\, {\dot x}^{j}\,-\,h( x^j + \theta^{jl} k_l , k_j )\,+ x^j\, V_j\,+\,k_j\, U^j\right]\right\}\,,
\eea

\noindent
where

\bml
\label{IV-7}
\bea
&& V_j\,\equiv\,J_j\,,\label{mlett:aIV-7}\\
&& U^j\,\equiv\,S^j\,-\,\theta^{j\,k}\,J_k\,.\label{mlett:bIV-7}
\eea
\eml

\noindent
Here, the dependence of $Z$ on the boundary values of $x$ and $t$ has been made explicit.

What remains to be elucidated is whether the path integral and the operator approaches yield equivalent descriptions for the quantum dynamics. We shall substantiate this proof of equivalence by reconstructing the equal time commutation relations in Eq.(\ref{1}) from the path integral approach. Since the equal time commutation relations are not modified by the interaction we may set, without loosing generality, $h( x^j + \theta^{jl} k_l , k_j )$ equal to the free Hamiltonian, namely,

\be
\label{IV-8}
h( x^j + \theta^{jl} k_l , k_j )\,=\,\frac{1}{2 M}\,k_j k_j\,.
\ee

\noindent
The path integral in Eq.(\ref{IV-6}) can now be performed explicitly and yields

\bea
\label{IV-9}
Z[J , K|x_f=0,t_f;x_{in}=0,t_{in}]\,=\,{\mathcal C}'\,\left(\det \Omega \right)^{- \frac{1}{2}}\,\exp\left\{\frac{i M}{2 \hbar}\,\int_{t_{in}}^{t_f}dt\,U^j(t) U^j(t)\right.\nonumber\\
\left.\frac{- i}{2 \hbar}\,\int_{t_{in}}^{t_f}dt\,\int_{t_{in}}^{t_f}dt'\,\left[V_j(t)-M {\dot U}^j(t)\right]\,\Delta^{jl}_F(t , t')\,\left[V_j(t')-M {\dot U}^j(t')\right]\right\}\,,
\eea

\noindent
where $\Omega_{jl}(t , t')$ is the local operator

\be
\label{IV-10}
\Omega_{jl}(t , t')\,=\,-\,M\,\delta_{jl}\,\frac{d^2 \delta(t - t')}{dt^2}\,,
\ee

\noindent
whose correspondent Green function ($\Delta^{jl}_F(t , t')$) is readily found to be

\be
\label{IV-11}
\Delta^{jl}_F(t , t')\,=\,\delta^{jl}\,\Delta_F(t , t')\,,
\ee

\noindent
with

\bea
\label{IV-12}
\Delta_F(t , t')\,&=&\,\frac{1}{M(t_f - t_{in})}\,\left[\theta(t - t')\,(t' - t_{in})\,(t_f - t)\right.\nonumber\\
&+&\,\left.\theta(t' - t)\,(t - t_{in})\,(t_f - t')\right]\,.
\eea

\noindent
Also, ${\mathcal C}'$ and $\det \Omega$ are constants.

We shall denote by $W[J , S|x_f,t_f;x_{in},t_{in}]$,

\be
\label{IV-13}
W[J , S|x_f,t_f;x_{in},t_{in}]\,\equiv\,\ln Z[J , S|x_f,t_f;x_{in},t_{in}]\,,
\ee

\noindent
the generating functional of normalized connected Green functions and by $T$ the chronological time ordering operator. Then, after some algebra the following two point Green functions are found

\bml
\label{IV-14}
\bea
&&\langle {E_0, t_f}|T \left(Q^l(t) Q^j(t')\right)|{E_0 , t_{in}}\rangle\,\equiv\,\left.\left(\frac{\hbar}{i}\right)^2\,\frac{\delta^2 W[J , S|x_f=0,t_f;x_{in}=0,t_{in}]}{\delta J_l(t) \delta J_j(t')}\right|_{J = S = 0}\nonumber\\
&&= i \hbar \delta^{lj}\,\Delta_F(t , t')\,+\,i \hbar \,\theta^{lj}\,\frac{(t - t')}{(t_f - t_{in})}\,-\,i \hbar \,\theta^{lj}\,\epsilon(t - t')\nonumber\\
&&-\,i \hbar \,M\,\left(\theta^2\right)^{lj}\,\delta(t - t')\,+\,i \hbar \,M\, \,\left(\theta^2\right)^{lj}\,\frac{1}{(t_f - t_{in})}\,,\label{mlett:aIV-14}\\
&&\langle{E_0, t_f}|T \left(Q^l(t) P_j(t')\right)|{E_0 , t_{in}}\rangle\,\equiv\,\left.\left(\frac{\hbar}{i}\right)^2\,\frac{\delta^2 W[J , S|x_f=0,t_f;x_{in}=0,t_{in}]}{\delta J_l(t) \delta S^j(t')}\right|_{J = S = 0}\nonumber\\
&&= i\, \hbar\, \delta^{l}_{\,\,j}\,M\,\frac{d\Delta_F(t , t')}{d t'}\,-\,i \hbar \,M\, \,\theta^{lj}\,\delta(t - t')\,+\,i \hbar \,M^2\,\theta^{lj}\,\frac{d^2\Delta_F(t , t')}{d t\, dt'}\,,\label{mlett:bIV-14}\\
&&\langle{E_0, t_f}|T \left(P_l(t) P_j(t')\right)|{E_0 , t_{in}}\rangle\,\equiv\,\left.\left(\frac{\hbar}{i}\right)^2\,\frac{\delta^2 W[J , S|x_f=0,t_f;x_{in}=0,t_{in}]}{\delta S^l(t) \delta S^j(t')}\right|_{J = S = 0}\nonumber\\
&&-\, i \hbar \delta_{lj}\,M\,\delta(t - t')\,+\, i \hbar M^2 \delta_{lj}\,\frac{d^2\Delta_F(t , t')}{d t\, dt'}\,,\label{mlett:cIV-14}
\eea
\eml

\noindent
where $|{E_0 , t}\rangle$ is the ground state energy eigenvector of the Heisenberg picture, $\epsilon(t)$ is the sign function and $t_{in} \leq (t , t') \leq t_f$. Now, the equal time commutator of any two operators, $A(t)$ and $B(t)$, say, can be expressed in terms of their chronological product ($T(A(t) B(t'))$) as follows

\be
\label{IV-15}
\left[A(t)\,,\,B(t)\right]\,=\,T(A(t) B(t'))\biggr|_{t = t'_{+}}\,-\,T(A(t) B(t'))\biggr|_{t = t'_{-}}\,,
\ee

\noindent
which clearly signalizes that the contributions to the commutator arise from the discontinuities of the chronological product at $t = t'$. Thus, the equal time commutator of two coordinate operators is only contributed by the term in the right hand side of Eq.(\ref{mlett:aIV-14}) containing the sign function. One finds

\be
\label{IV-16}
\langle{E_0, t_f}|\left[Q^l(t)\,,\, Q^j(t)\right]|{E_0 , t_{in}}\rangle\,=\,-\,2 i \hbar\,\theta^{l\,j}\,.
\ee

\noindent
The contribution to the equal time commutator $\left[Q^l(t)\,,\, P^k(t)\right]$ arises from the discontinuity at $t = t'$ exhibited by the first term in the right hand side of Eq.(\ref{mlett:bIV-14}) and, hence,

\be
\label{IV-17}
\langle{E_0, t_f}|\left[Q^l(t)\,,\, P_j(t)\right]|{E_0 , t_{in}}\rangle\,=\, i \hbar\,\delta^{l}_{\,\,j}\,.
\ee

\noindent
Finally, the fact that all terms in the right hand side of Eq.(\ref{mlett:cIV-14}) are continuous at $t = t'$ leads to

\be
\label{IV-18}
\langle{E_0, t_f}|\left[P_l(t)\,,\, P_j(t)\right]|{E_0 , t_{in}}\rangle\,=\,0\,.
\ee

\noindent
It is obvious that the matrix elements of the basic commutators arising from the functional approach are in agreement with the commutation rules in Eq.(1). We, then, conclude that the operator and the functional frameworks provide equivalent descriptions of the quantum dynamics for noncommutative models.

\section{The Feynman kernel of the two dimensional noncommutative harmonic oscillator}
\label{sec:level5}

Our purposes in this section is to exhibit the intricacies arising along the computation of the Feynman kernel for a noncommutative system. For reasons of feasibility we shall be dealing here with an exactly solvable model: the noncommutative two dimensional harmonic oscillator\cite{Dragovich1}. This will also allow us to discuss the similarities and discrepancies existing between the commutative and noncommutative versions of the theory.

Hence, for the model under scrutiny

\be
\label{3-1}
h(q , p)\,=\,\frac{p^j p^j}{2 M}\,+\,\frac{\omega^2}{2}\,q^j q^j\,,
\ee

\noindent
where $M$ and $\omega$ are, respectively, the mass and the frequency of the oscillator, while repeated spatial indices only sum from 1 to 2. Correspondingly, the Hamiltonian $h( x^j + \theta^{jl} k_l , k_j )$ reads

\be
\label{3-2}
h( x^j + \theta^{jl} k_l , k_j )\,=\,\frac{k_j k_j}{2M}\,+\,\frac{M\omega^2}{2}\left(x^j x^j +2x^i\theta^{ij}k_j+\theta^{ij}\theta^{il}k_j k_l\right)\,.
\ee

Then, as is well known, the Feynman kernel ($K(x_f , t_f ; x_{in} , t_{in})$) is given by the phase space path integral

\be
\label{3-3}
K(x_f , t_f ; x_{in} , t_{in})\,=\,\int {\mathcal D}x\,{\mathcal D}k\,\exp\left\{\frac{i}{\hbar}\int_{t_{in}}^{t_f} dt
\left[k_j\,{\dot x}^j\,-\,h( x^j + \theta^{jl} k_l , k_j )\right]\right\}\,,
\ee

\noindent
where $x_{in}$ ($x_f$) denote, as we already said, the values acquired by the coordinates at $t = t_{in}$ ($t = t_f$). After carrying out the momentum integrals one arrives at

\be
\label{3-4}
K(x_f , t_f ; x_{in} , t_{in})\,=\,{\mathbb C}\,\int {\mathcal D}x\, e^{\frac{i}{\hbar}\,S[x]}\,,
\ee

\noindent
where ${\mathbb C}$ is a constant,

\be
\label{3-5}
S[x]\,=\,\int_{t_{in}}^{t_f} dt \,{\mathcal L}(x(t) , {\dot x}(t))
\ee

\noindent
denotes the effective action functional and

\bea
\label{3-6}
{\mathcal L}(x(t) , {\dot x}(t))\,=\,\frac{1}{2}\,M_{\theta}\,{\dot x}^j {\dot x}^j\,-\,M_{\theta}\,M\,\omega^2\,x^i\,\e^{ij}\,{\dot x}^j\,-\,\frac{1}{2}\,M_{\theta}\,\omega^2\,x^i\,x^i\,,
\eea

\noindent
is the effective Lagrangian. We furthermore recall that in two dimensions one can write $\theta^{jk} = \e^{jk} \,\theta$, where $\e^{jk}$ is the antisymmetric Levi-Civita tensor and $\theta$ is a scalar parameterizing the intensity of the noncommutativity. Also, we have introduced the definition

\be
\label{3-7}
M_{\theta}\,\equiv\,\frac{M}{1 + M^2 \omega^2 \theta^2}\,.
\ee

As it can be seen, the Lagrangian in Eq.(\ref{3-6}) is bilinear in $x$. Therefore, the functional integral in the right hand side of Eq.(\ref{3-4}) can also be exactly computed and yields

\be
\label{3-8}
K(x_f , t_f ; x_{in} , t_{in})\,=\,{\mathcal N}\,e^{\frac{i}{\hbar}\,S[x_{cl}]}\,,
\ee

\noindent
where ${\mathcal N}$ is another constant and $x_{cl}$ are the solutions of the Lagrange equations of motion deriving from (\ref{3-6}), namely,

\bml
\label{3-9}
\bea
&&{\ddot x}^1\,+\,2\,\theta\,M\,\omega^2\,{\dot x} ^2\,+\,\omega^2\,x^1\,=\,0\,,\label{mlett:a3-9}\\
&&{\ddot x}^2\,-\,2\,\theta\,M\,\omega^2\,{\dot x} ^1\,+\,\omega^2\,x^2\,=\,0\,.\label{mlett:b3-9}
\eea
\eml

\noindent
One can convince oneself that

\be
\label{3-10}
S[x_{cl}]\,=\,\frac{1}{2}\,M_{\theta}\,\left[ x^j(t_f)\,{\dot x}^j(t_f)\,-\, x^j(t_{in})\,{\dot x}^j(t_{in})\right]\,.
\ee

What remains to be done is to find the configurations $x^j(t)$ solving the coupled ordinary differential equations of motion (\ref{3-9}) under the boundary conditions $x^j=x^j_{in}$, for $t=t_{in}$, and $x^j=x^j_f$, for $t = t_f$. The corresponding decoupling is easily implemented by introducing the chiral variable $z \equiv(x^1 + i \, x^2)/\sqrt{2}$.
One, then, finds

\bml
\label{3-11}
\bea
x^1(t)\,&=&\,\frac{1}{\sin\left[\omega\sqrt{\kappa}(t_f-t_{in})\right]}\,\left\{x^1_{in}\,
\sin\left[\omega\sqrt{\kappa}(t_f-t)\right]\cos\left[M\theta\omega^2(t-t_{in})\right]\right.\nonumber\\
&-&\,
x^2_{in}\,\sin\left[\omega\sqrt{\kappa}(t_f-t)\right]\sin\left[M\theta\omega^2(t-t_{in})\right]\nonumber\\
&+&\,
x^1_f\,\sin\left[\omega\sqrt{\kappa}(t-t_{in})\right]\cos\left[M\theta\omega^2(t_f-t)\right]\nonumber\\
&+&\left.\,
x^2_f\,\sin\left[\omega\sqrt{\kappa}(t-t_{in})\right]\sin\left[M\theta\omega^2(t_f-t)\right]\right\}\,,
\label{mlett:a3-11}\\
x^2(t)\,&=&\,\frac{1}{\sin\left[\omega\sqrt{\kappa}(t_f-t_{in})\right]}\,\left\{x^1_{in}\,
\sin\left[\omega\sqrt{\kappa}(t_f-t)\right]\sin\left[M\theta\omega^2(t-t_{in})\right]\right.\nonumber\\
&+&\,
x^2_{in}\,\sin\left[\omega\sqrt{\kappa}(t_f-t)\right]\cos\left[M\theta\omega^2(t-t_{in})\right]\nonumber\\
&-&\,
x^1_f\,\sin\left[\omega\sqrt{\kappa}(t-t_{in})\right]\sin\left[M\theta\omega^2(t_f-t)\right]\nonumber\\
&+&\left.\,
x^2_f\,\sin\left[\omega\sqrt{\kappa}(t-t_{in})\right]\cos\left[M\theta\omega^2(t_f-t)\right]\right\}\,,
\label{mlett:b3-11}
\eea
\eml

\noindent
where

\be
\label{3-12}
\kappa\equiv 1+M^2\theta^2\omega^2\,.
\ee

\noindent
By substituting Eq.(\ref{3-11}) into (\ref{3-10}) one arrives at

\bea
\label{3-13}
S[x_{cl}]\,&=&\,\frac{M_{\theta}}{2}\frac{\omega\sqrt{\kappa}}{\sin\left[\omega\sqrt{\kappa}(t_f-t_{in})\right]}
\left\{\cos\left[\omega\sqrt{\kappa}(t_f-t_{in})\right]\left(x^j_fx^j_f+x^j_{in}x^j_{in}\right)\right.\nonumber\\
&-&\left.
2\cos\left[M\theta\omega^2(t_f-t_{in})\right]x^j_{in}x^j_{f}
+2\sin\left[M\theta\omega^2(t_f-t_{in})\right]\e^{jk}x^j_{f}x^k_{in}\right\}\,,
\eea

\noindent
which at the commutative limit ($\theta=0$) reduces, as expected, to two uncoupled harmonic oscillators\cite{Feynman}.

The last term within the curly bracket, in the right hand side of Eq.(\ref{3-13}), describes the most striking effect introduced by the noncommutativity. It shows that, for $\theta \neq 0$, the coordinates $x^1$ and $x^2$ become mixed. The same effect will of course occur in the case of the Landau problem since this last mentioned system can be fully rephrased in terms of a noncommutative two dimensional harmonic oscillator\cite{Girotti1}.

\section{Conclusions and final remarks}
\label{sec:level6}

This work was primarily intended to review model independent results in noncommutative quantum mechanics.

We first presented a unified description of the classical and quantum dynamics of a generic noncommutative system. The distinctive feature is that at the classical level one deals with a constrained system whose quantization leads to the non-canonical commutation rules in Eq.(\ref{1}), which act as input in the formulation of the problem. We can not assert that the classical-quantum correspondence depicted in Section 2 is unique, since one can not rule out the possibility of existing another constrained system whose DB's are still those given in Eq.(\ref{5}). However, the physical variables for this new system may, at the most, differ from $x$, $k$ by a canonical transformation.

The fact that non-commutativity does not destroy the Born series greatly facilitated the proof of unitarity, which is an essential requirement for a quantum theory to make sense.

The work done in Refs.\cite{Dirac1,Fradkin1,Sundermeyer1,Girotti2,Gitman1,Sudarshan1}, in connection with constrained systems, paved the way for us to implement the functional formulation of the quantum dynamics of noncommutative systems. We succeeded in recovering the input information in Eq.(\ref{1}) from the functional formalism. As shown in Section 5, in connection with the noncommutative two dimensional harmonic oscillator, the main effect induced by the noncommutativity consists in mixing the degrees of freedom of the physical system.

Both of us acknowledge partial support from Conselho Nacional de Desenvolvimento Cient\'{\i}fico e Tecnol\'ogico (CNPq), Brazil.


\begin{thebibliography}{99}
\bibitem {footnote 0} Internal degrees of freedom such as spin, isospin, etc., do not bring additional difficulties and are, therefore, left out of consideration.
\bibitem {footnote1} Observe that the dimension of $\theta^{lk}$ in the cgs system is $d[\theta] =$ g$^{-1}$ s.
\bibitem {footnote101} There is a plethora of references concerned with non-commutative quantum mechanics. However, we only mention here our papers of principle.
\bibitem {footnote3} Everywhere in this paper $q$-numbers are denoted by capital letters while $c$-numbers are denoted by lower case letters.
\end{thebibliography}
\end{document}